\documentclass[twocolumn]{aastex6}
\usepackage{apjfonts}

\newcommand{\Ka}{\ion{Fe}{1}~K$\alpha$}
\newcommand{\Hea}{\ion{Fe}{25}~He$\alpha$}
\newcommand{\Lya}{\ion{Fe}{26}~Ly$\alpha$}
\newcommand{\Kb}{\ion{Fe}{1}~K$\beta$}
\newcommand{\Heb}{\ion{Fe}{25}~He$\beta$}
\newcommand{\Lyb}{\ion{Fe}{26}~Ly$\beta$}
\newcommand{\NiKa}{\ion{Ni}{1}~K$\alpha$}
\newcommand{\NiHea}{\ion{Ni}{27}~He$\alpha$}

\begin{document}


\title{Evidence for a neutral iron line generated by MeV protons from supernova remnants interacting with molecular clouds}

%% Use \author, \affil, plus the \and command to format author and affiliation 
%% information.  If done correctly the peer review system will be able to
%% automatically put the author and affiliation information from the manuscript
%% and save the corresponding author the trouble of entering it by hand.
%%
%% The \affil should be used to document primary affiliations and the
%% \altaffil should be used for secondary affiliations, titles, or email.

%% Authors with the same affiliation can be grouped in a single
%\author{Kumiko K. Nobukawa\altaffilmark{1}, et al.}
\author{Kumiko K. Nobukawa\altaffilmark{1}, Masayoshi Nobukawa\altaffilmark{2}, Katsuji Koyama\altaffilmark{3}, Shigeo Yamauchi\altaffilmark{1}, Hideki Uchiyama\altaffilmark{4}, Hiromichi Okon\altaffilmark{3}, Takaaki Tanaka\altaffilmark{3}, Hiroyuki Uchida\altaffilmark{3}, and Takeshi G. Tsuru\altaffilmark{3}}
\altaffiltext{1}{Department of Physics, Faculty of Science, Nara Women's University, Kitauoyanishi-machi, Nara, Nara 630-8506, Japan}
\altaffiltext{2}{Faculty of Education, Nara University of Education, Takabatake-cho, Nara, Nara 630-8528, Japan}
\altaffiltext{3}{Department of Physics, Graduate School of Science, Kyoto University, Kitashirakawa-oiwake-cho, Sakyo-ku, Kyoto, Kyoto 606-8502, Japan}
\altaffiltext{4}{Faculty of Education, Shizuoka University, 836 Ohya, Suruga-ku, Shizuoka, Shizuoka 422-8529, Japan}
%% Use the \and command so offset the last author.

\email{kumiko@cc.nara-wu.ac.jp}


%% Mark off the abstract in the ''abstract'' environment. 
\begin{abstract}
Supernova remnants (SNRs) have been prime candidates for Galactic cosmic-ray accelerators.  When low-energy cosmic-ray protons  (LECRp) collide with interstellar gas, they ionize neutral iron atoms and emit the neutral iron line (\Ka) at 6.40~keV. We search for the iron K-shell line in seven SNRs from the {\it Suzaku} archive data of the Galactic plane in the  $6\degr \lesssim l \lesssim 40\degr, |b| < 1\degr$ region.
All these SNRs interact with molecular clouds.
We discover \Ka\ line emissions from five SNRs (W28, Kes\,67, Kes\,69, Kes\,78, and W44).
The spectra and morphologies suggest that the \Ka\ line is produced by interactions between LECRp and the adjacent cold gas. The proton energy density is estimated to be $\gtrsim$~10--100 eV~cm$^{-3}$, which is more than 10 times higher than that in the ambient interstellar medium.  

\end{abstract}

\keywords{cosmic rays --- X-rays: ISM --- ISM: supernova remnants}



\section{Introduction} \label{sec:intro}
Supernova remnants (SNRs) are thought to produce Galactic cosmic rays (CRs) via diffusive shock acceleration (DSA).  
Observational evidence for cosmic-ray acceleration came from {\it ASCA}, which detected the non-thermal X-ray emission due to electrons accelerated to the TeV energy band \citep{1995Natur.378..255K}.  Recently GeV and TeV gamma-ray observations have revealed that protons and/or electrons are accelerated to energies up to 100~TeV in SNR shells  \citep[e.g.,][]{2013Sci...339..807A, 2007A&A...464..235A}. 
In the hadronic origin, the GeV and TeV gamma-rays come from decays of $\pi^{0}$ particles, which are produced by  interactions between interstellar gas and high-energy cosmic-ray protons above the threshold energy of  280~MeV.  

In DSA, suprathermal particles (low-energy cosmic rays; LECRs) are accelerated to relativistic energy  
through multiple crossings of shock front \citep[e.g.,][]{2001RPPh...64..429M}. 
Thus, the energy spectra and fluxes of the LECRs in SNRs provide a key link 
to generation of the GeV and TeV CRs in SNRs. 
By extrapolating the spectra of high-energy protons obtained by gamma-ray observations \citep{2013Sci...339..807A} to the low-energy end, we estimate that the (kinetic) energy density of MeV protons typically amounts to about 1--10~eV~cm$^{-3}$.
However, there has been very  few observation of  LECRs below the MeV band due to lack of an effective probe to investigate them. 

Observations of $\rm{H}^{+}_3$  absorption lines generated by ionization of $\rm{H}_2$ molecules have provided information on the cosmic-ray ionization rate \citep[e.g.,][]{2012ApJ...745...91I}.
While the cross section for $\rm{H}_2$ ionization by protons peaks at the keV band,  in a dense cloud with the typical column density of molecular hydrogen $N({\rm H}_2)=10^{22}$~cm$^{-2}$, the bulk of the ionization is dominated by CRs in the MeV--GeV band and followed by those in  the keV band \citep{2009A&A...501..619P}. 
Although $\rm{H}^{+}_3$ is produced anywhere with $\rm{H}_2$ molecules, measurements of the ionization rate require bright infrared sources as a background, and thus there are limited samples of $\rm{H}^{+}_3$ observations yet \citep{2012ApJ...745...91I}. Furthermore,   $\rm{H}^{+}_3$ observations  cannot provide the information on a spectrum and energy density of LECRs.  

Voyager-I  traveled  122 astronomical units away from the Earth and observed the spectra of CRs immediately outside the heliosphere down to several MeV per nucleon without an effect by solar modulation \citep{2013Sci...341..150S}.   
Although Voyager-I advances our understanding of LECRs,  in order to measure their spectra and energy density in SNRs or the Galactic interstellar space, we should observe LECRs just near the acceleration source because of the short diffusion length of LECRs.  

\cite{2012A&A...546A..88T} studied signatures that allow  identification of radiation from LECRs in X-ray spectra showing the neutral iron line (\Ka) at 6.40~keV. By applying the developed models to the X-ray emission from the Arches cluster region near the Galactic center, the authors found that the \Ka\ line from the region is likely produced by LECRs.
\cite{2015ApJ...807L..10N} have also demonstrated that the \Ka\ line can be a unique and powerful probe to investigate LECR protons (LECRp) in the observations of the Galactic disk at $l = 1{\degr}.5$--$3{\degr}.5$. When protons in the MeV band collide with molecular clouds (MCs), the \Ka\  line is produced via inner-shell ionization of neutral iron.  As for SNRs, there are only two possible samples where the LECR-induced \Ka\ line was detected: \cite{2014PASJ...66..124S, 2016PASJ...68S...8S} found from two SNRs (3C\,391 and Kes\,79) a hint of the \Ka\ line feature, which can be produced by interactions of locally accelerated protons with the surrounding MCs.  
All the above \Ka\ line emissions but for the Arches cluster were detected by sensitive observations with the X-ray Imaging Spectrometer (XIS) onboard \textit{Suzaku}, which has advantages of the low and stable background and high sensitivity, especially in the iron K-shell band \citep{Mitsuda:2007km, Koyama:2007wg}.   
  
In order to investigate the \Ka\ line in SNRs in our galaxy, we should carefully estimate the X-ray background, which is dominated by the Galactic ridge X-ray emission (GRXE). The distribution and spectrum of the GRXE is well studied in the inner Galactic ridge \citep[$|l|<30\degr-40\degr$,  $|b| < 1\degr$;][]{2016ApJ...833..268N, 2016PASJ...68...59Y}.  
The \Ka\ intensity smoothly distributes in the eastern side although there seems to be a local structure in the western side (figure 2 of \citealt{2016PASJ...68...59Y}). 
In this paper, we focus on the eastern region of the inner Galactic ridge, and
 search for the \Ka\  line in seven SNRs from  the archives of the {\it Suzaku} XIS.  
We then newly detect the \Ka\ line from five of the SNRs. 
Based on the spatial and spectral analysis, we discuss the origin of the \Ka\ line.  The error bars given in the spectra are at the 1$\sigma$ confidence level. 

\floattable
\begin{deluxetable}{lcrrrrrrr}
\tablecaption{Observation log and  intensities of the \Ka\ and \Hea\ lines and the line ratio for each FOV. 
\label{tab:FOV}}
\tablecolumns{9}
\tablenum{1}
\tablewidth{0pt}
\tablehead{
\colhead{Object} &
\colhead{Obs. ID} &
\multicolumn2c{Pointing direction} &
\colhead{Obs. start } &
\colhead{Exposure} &
\multicolumn2c{Line intensity\tablenotemark{a,b}} &
\colhead{Ratio\tablenotemark{b,c}} \\
\colhead{} &
\colhead{} &
\colhead{$l(\degr)$} &  \colhead{$b(\degr$)} &
\colhead{(UT)} & 
\colhead{(ks)} &
\colhead{\Ka\ } &
\colhead{\Hea\ } &
\colhead{} 
}
\startdata
W28 		& 505005010 	& $6.47$ & $-0.00$	& 2010 Apr 03 & 73.0	& 2.41$\pm$0.43	& 11.76$\pm$0.55	& 0.21$\pm$0.04  \\
		& 505006010 	& $6.68$ & $-0.20$	& 2011 Feb 25 & 100.0	& 2.95$\pm$0.38	& 6.82$\pm$0.42	& 0.43$\pm$0.06 \\
		& 506036010 	& $6.15$ & $0.07$	& 2011 Oct 10 & 151.1 	& 2.29$\pm$0.32		& 12.86$\pm$0.42	& 0.18$\pm$0.03 \\
		& 508006010 	& $6.38$ & $-0.24$	& 2014 Mar 22 & 40.9	& 3.21$\pm$0.64	& 8.10$\pm$0.75	& 0.40$\pm$0.09 \\
		& 508006020 	& $6.39$ & $-0.25$	& 2014 Oct 08 & 61.7	& 3.36$\pm$0.50	& 8.22$\pm$0.57	& 0.41$\pm$0.07 \\ \hline
Kes\,67	& 506051010 	& $18.78$ & $0.40$	& 2012 Mar 08 & 52.0	& 2.74$\pm$0.54	& 5.98$\pm$0.60	& 0.46$\pm$0.10 \\ \hline
Kes\,69	& 509037010 	& $21.82$ & $-0.57$ & 2014 Sep 27 & 77.3	& 3.55$\pm$0.42	& 5.17$\pm$0.44	& 0.69$\pm$0.10 \\ \hline
Kes\,75	& 404081010 	& $29.77$ & $-0.20$ & 2009 Apr 15  & 104.3	& 1.33$\pm$0.56	& 5.91$\pm$0.61	& 0.22$\pm$0.10 \\ \hline
Kes\,78	& 507035010 	& $32.88$ & $-0.00$ & 2012 Apr 20 & 55.1	& 2.56$\pm$0.64	& 4.82$\pm$0.68	& 0.53$\pm$0.15 \\
		& 507036010 	& $32.69$ & $-0.08$ & 2012 Apr 21 & 52.2	& 1.15$\pm$0.45	& 5.76$\pm$0.52	& 0.20$\pm$0.09 \\ \hline
W44		& 505004010 	& $34.70$ & $-0.41$ & 2010 Apr 10 & 61.2	& 2.37$\pm$0.47	& 3.51$\pm$0.49	& 0.68$\pm$0.16 \\
		& 508002010 	& $34.60$ & $-0.36$ & 2013 Oct 24 & 61.1	& 2.00$\pm$0.53	& 3.47$\pm$0.57	& 0.57$\pm$0.18 \\		
		& 508003010 	& $34.56$ & $-0.52$ & 2013 Oct 22 & 66.7	& 1.95$\pm$0.54	& 3.26$\pm$0.54 	& 0.60$\pm$0.19 \\		
		& 508003020 	& $34.56$ & $-0.51$ & 2014 Apr 09 & 32.4	& 1.61$\pm$0.82	& 3.69$\pm$0.83 	& 0.44$\pm$0.24 \\		
		& 508004010 	& $34.67$ & $-0.55$ & 2013 Oct 18 & 58.3	& 1.23$\pm$0.44	& 4.02$\pm$0.51	& 0.31$\pm$0.11 \\		
		& 508005010 	& $34.76$ & $-0.30$ & 2013 Oct 19 & 55.7	& 1.48$\pm$0.49	& 4.27$\pm$0.52	& 0.35$\pm$0.12  \\ \hline
3C\,396 	& 509038010 	& $39.19$ & $-0.30$ & 2014 Apr 26 & 82.8	& 1.45$\pm$0.39	& 5.30$\pm$0.44	& 0.27$\pm$0.08 \\		
\enddata
\tablenotetext{a}{Units are $10^{-8}$~photons~cm$^{-2}$~s$^{-1}$~arcmin$^{-2}$.}
\tablenotetext{b}{Errors are quoted at the 1$\sigma$ level.}
\tablenotetext{c}{Intensity ratio of \Ka/\Hea.}
\end{deluxetable}

\section{Observations and Data reduction} \label{sec:obs}
In the archive data of the {\it Suzaku} XIS, we picked up seven X-ray emitting SNRs 
which are located in the region of 
$6\degr \lesssim l \lesssim 40\degr, |b| < 1\degr$: W28, Kes\,67, Kes\,69, Kes\,75, Kes\,78, 3C\,396, and W44.  
The observation and analysis log is listed in table~\ref{tab:FOV}. 
Three out of the seven samples, W28, Kes\,69, and W44, are classified as mixed-morphology SNRs \citep{1998ApJ...503L.167R, 2010ApJ...712.1147J}, associated with MCs.
Accelerated cosmic-ray particles can bombard the MCs to produce the \Ka\ line.

The XIS consists of four X-ray CCD cameras placed on the focal plane of the X-Ray Telescope \citep[XRT;][]{2007PASJ...59S...9S}. The XIS has a field of view (FOV) of $17.'8 \times17.'8$. Three of the sensors (XIS\,0, 2, and 3) have front-illuminated (FI) devices, while the other one (XIS\,1) contains a back-illuminated (BI) device. The whole FOV of XIS\,2 and one-fourth of XIS\,0 have been out of function since 2006 November and 2009 June, respectively, and thus the data from the remaining devices were used. 

We reprocessed the data by using \texttt{xispi} in the analysis software package, HEAsoft 6.18.1, and the {\it Suzaku} calibration database (CALDB) released in 2016 February,  with the standard event selection criteria for the XIS data processing. The response file (arf) and redistribution file (rmf)  were produced by \texttt{xissimarfgen} and \texttt{xisrmfgen} \citep{2007PASJ...59S.113I}, respectively.
The non-X-ray background (NXB) was estimated by \texttt{xisnxbgen} \citep{2008PASJ...60S..11T}.

\section{Analysis and Results} \label{sec:results}

\subsection{The iron K-shell line flux from the XIS field} \label{ironline}
We made an X-ray spectrum from each FOV of the SNR observations (table~\ref{tab:FOV}), excluding point-like sources in the fields. 
X-ray spectra of all the picked-up SNRs but for Kes\,67 have been well studied:  the electron temperature of SNR plasma is  $\lesssim 1$~keV 
\citep{2012PASJ...64...81S, 2012A&A...541A.152B, 2016ApJ...818...63B, 2009ApJ...694..376S, 2012PASJ...64..141U, 2011ApJ...727...43S}.
Since no previous report is available for Kes\,67, we analyzed the full-band spectrum and obtained the temperature of $0.4\pm0.1$~keV.
Such low-temperature plasmas are dim above 5~keV band and hardly emit iron K-shell lines. 

The seven SNRs are located on the Galactic ridge ($6{\degr}\lesssim l \lesssim 40{\degr}, |b|<1{\degr}$), and hence the X-ray flux above 5~keV is dominated by the GRXE \citep{2013PASJ...65...19U, 2016PASJ...68...59Y}, which has strong K-shell lines of neutral (\Ka), helium-like (\Hea), and hydrogen-like iron (\Lya) at 6.40~keV, 6.68 keV, and 6.97 keV, respectively. 
The only exception is the FOV containing Kes\,75,  which has a pulsar, PSR\,J1846$-$0258, and its pulsar wind nebula,  \citep[PWN;][]{2003ApJ...582..783H}. {\it Suzaku} XRT cannot spatially resolve the X-ray bright point-like source and the thermal component. Non-thermal emission from the source is not negligible in the 5--8 keV band. However, since the pulsar and PWN show no emission line,  it does not affect the iron K-shell line measurement. 

We subtracted the NXB and the cosmic X-ray background (CXB) from the spectra. The CXB is expressed as a power law with the photon index of 1.4 and the flux of $8.2\times10^{-7}$~photons~cm$^{-2}$~s$^{-1}$~arcmin$^{-2}$~keV$^{-1}$ at 1 keV \citep{2002PASJ...54..327K}. We fitted the 4--9 keV band spectra with a phenomenological model consisting of an absorbed power law and four Gaussians. The line energies of the Gaussians are fixed to 6.40, 6.68, 6.97, and 7.06 keV (Fe{\small I}-K$\beta$) \citep{1993A&AS...97..443K, Smith:2001er}. The intensity of the \Kb\ line is fixed to 0.125 times that of the \Ka\ line \citep{1993A&AS...97..443K}, while those of the other lines are free parameters. Since the absorption has no significant effect on the relevant energy band, absorption column density of the interstellar medium is fixed to $3\times10^{22}$~cm$^{-2}$ \citep{2013PASJ...65...19U}. 
We used bremsstrahlung instead of a power law, but the line intensities did not change.

The best-fit intensities of the \Ka\ and \Hea\ lines, and the intensity ratio of \Ka/\Hea\ in the individual FOVs are summarized in table~\ref{tab:FOV}.  The \Hea\ line intensity in W28 is the highest among the SNRs. This is because W28 is located closest to the Galactic center among the SNRs ($l\sim6{\degr}$, $b\sim-0.2{\degr}$), where the GRXE becomes stronger with decreasing distance from the 
Galactic center  \citep{2013PASJ...65...19U, 2016PASJ...68...59Y}.
The intensity ratios of \Ka/\Hea\ from all the fields of the seven SNRs are 
compared to the averaged one in the GRXE of  $0.27\pm0.02$ 
\citep[at the 1$\sigma$ level, see table~\ref{tab:integ},][]{2016ApJ...833..268N}.  
The flux ratios of \Ka/\Hea\ in the FOVs containing W28, Kes\,67, Kes\,69, Kes\,78, and W44 are higher than the average of the GRXE with the 1$\sigma$ level,  while those containing Kes\,75 and 3C\,396 are consistent with the GRXE within 1$\sigma$ errors.
Thus, we suspect that the five SNRs, W28, Kes\,67, Kes\,69, Kes\,78, and W44, have enhanced \Ka\ line emissions.
We examine the enhancement of the \Ka\ line emission in detail by spectral analyses in the next subsection.

\subsection{X-ray spectra from the \Ka-enhanced and reference regions} % sec 3.2
\label{sec:image_spec}

\begin{figure*}[tbh]
\figurenum{1}
\begin{center}
\includegraphics[width=12cm]{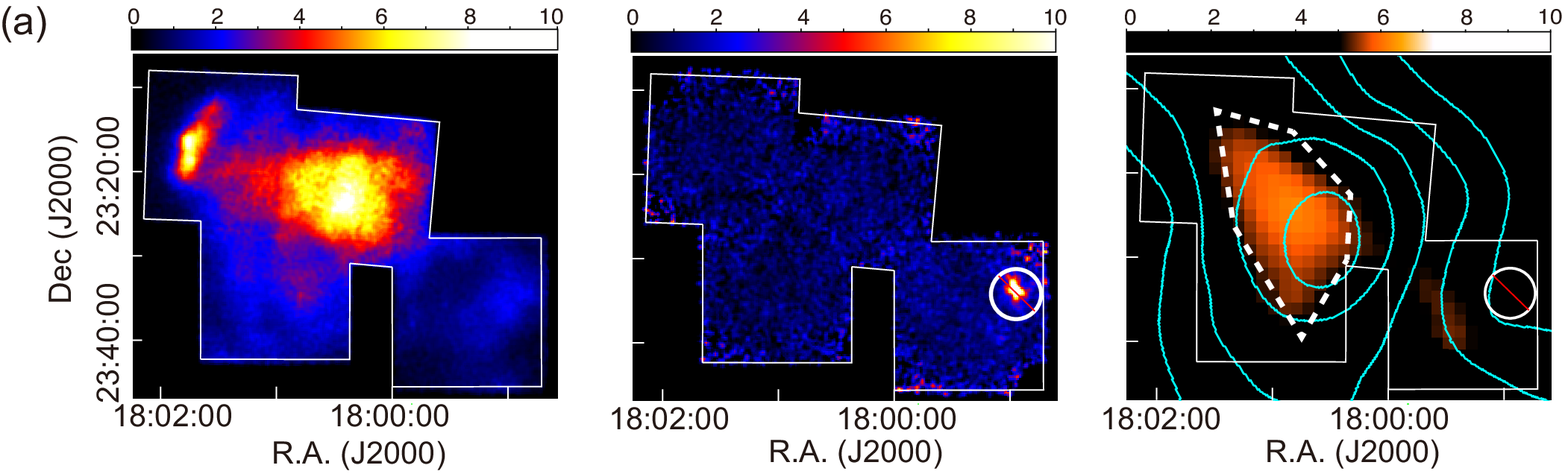}
\includegraphics[width=12cm]{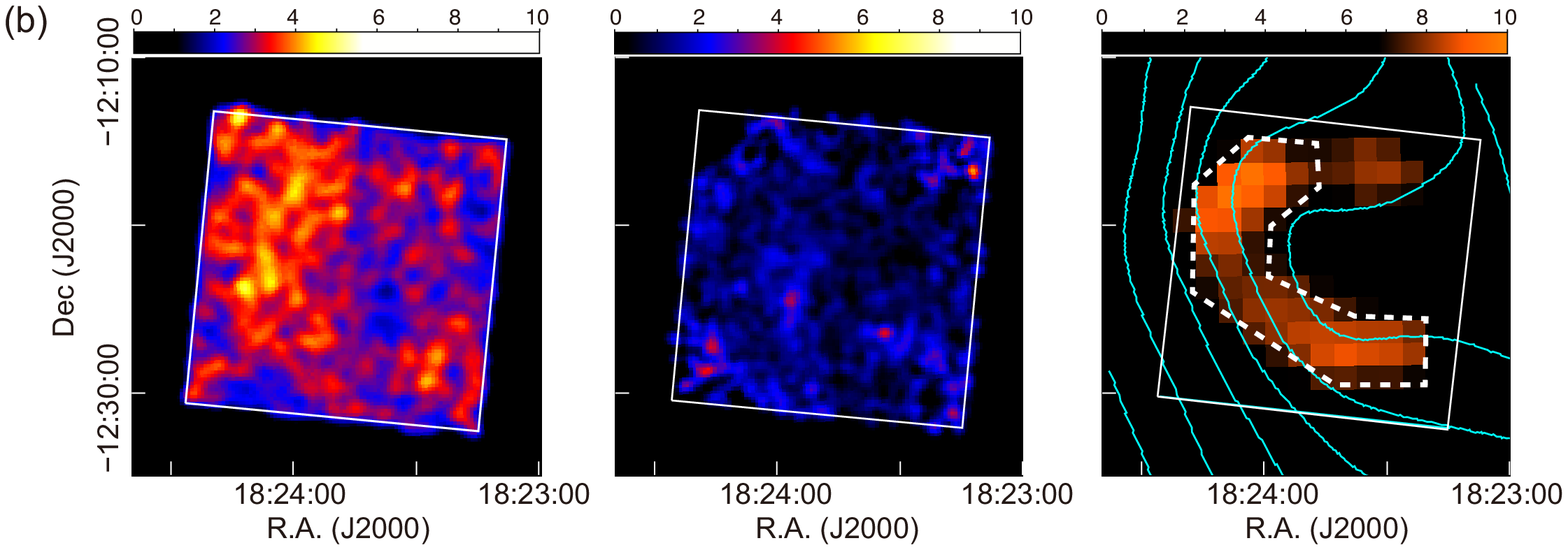}
\includegraphics[width=12cm]{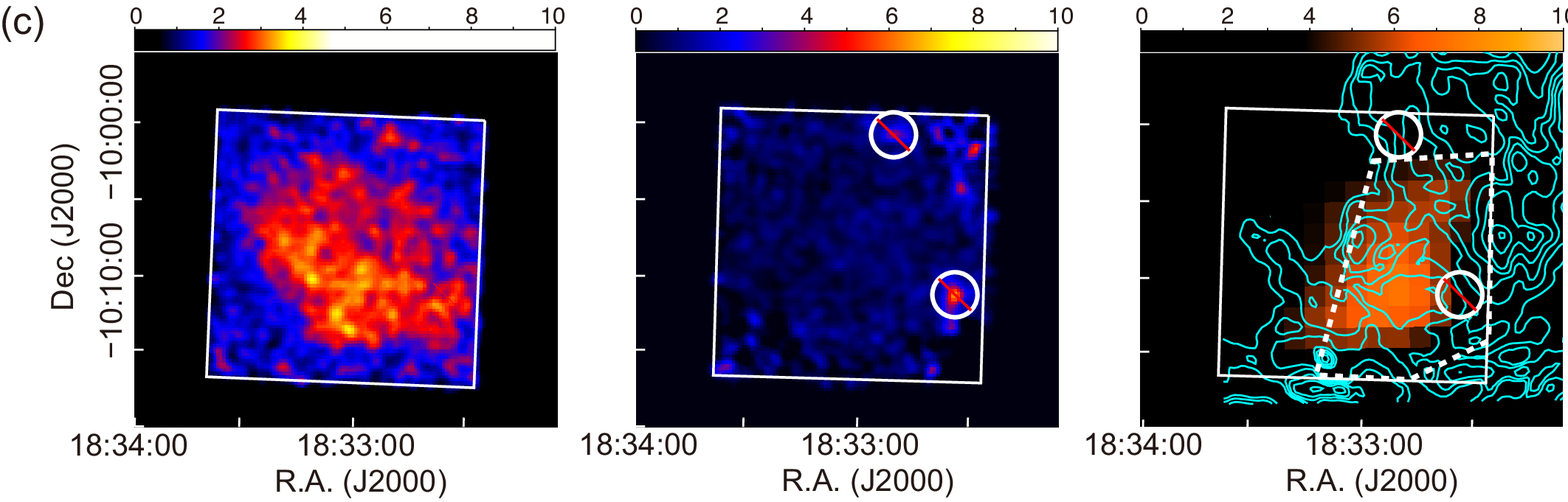}
\includegraphics[width=12cm]{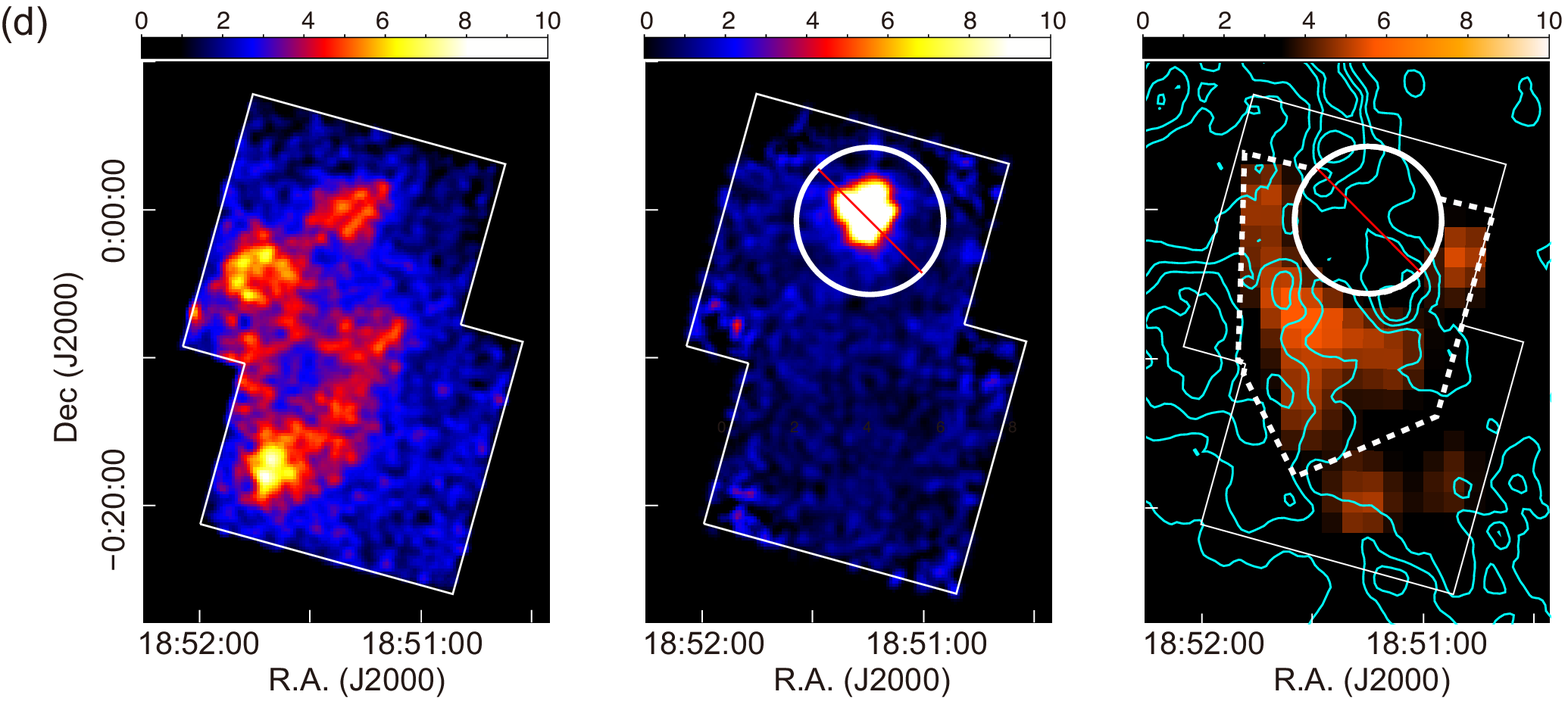}
\includegraphics[width=12cm]{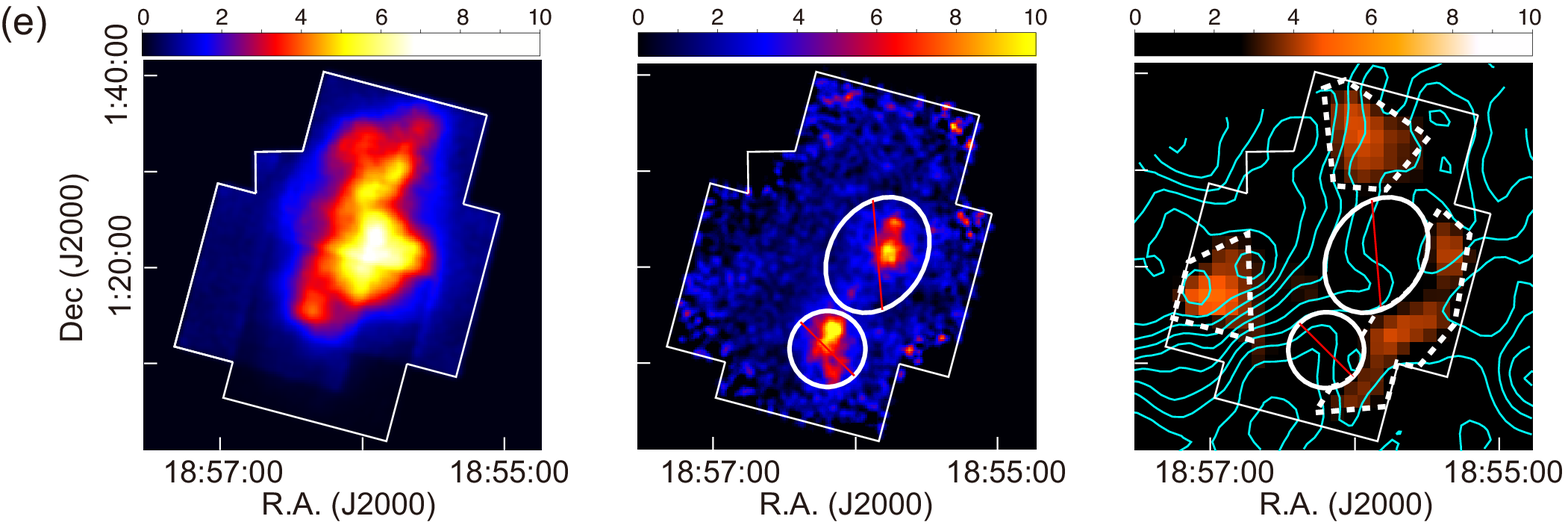}
\caption{Band images of (a) W28, (b) Kes\,67, (c) Kes\,69, (d) Kes\,78, and (e) W44. 
The left, center, and right  panels show the images of the soft (0.5--2~keV), hard (5--8~keV), and the \Ka\  line band (6.2--6.5~keV), respectively (color bars in arbitrary unit). Vignetting is corrected and the NXB is subtracted.
The regions surrounded by the white lines in the individual panels show the FOVs of the XIS. 
Point-like or slightly extended sources are marked with solid white circle or ellipse regions, which are described in section~\ref{sec:image_spec}. 
In the right panels, those regions are excluded from the images. 
The dashed lines in the right panels indicate the ``enhanced regions'' (see text).  
The contours in the right panels show the distributions of H{\small I} at 37.5~km~s$^{-1}$ \citep{2002AJ....124.2145V}, H{\small I} at $18.1$~km~s$^{-1}$ \citep[][the intensity of H{\small I} increases towards the east]{1999AJ....118..930D}, $^{12}$CO ($J=$1--0) at 80--81~km~s$^{-1}$  \citep{2009ApJ...691..516Z}, $^{13}$CO ($J=$1--0) at 80--84~km~s$^{-1}$ \citep{2011ApJ...743....4Z}, 
and  $^{12}$CO ($J=$2--1)   at 40.0--50.3~km~s$^{-1}$  \citep{2013ApJ...768..179Y}, for (a), (b), (c), (d), and (e), respectively. }
\label{fig:images}  
\end{center}
\end{figure*}

In order to investigate the \Ka\ line enhanced regions of the five SNRs, we made the \Ka\ line band images (6.2--6.5 keV) 
with the binning size of $80\times80$~pixels ($1.4\times1.4$~arcmin$^2$) and Gaussian smoothing is applied with $\sigma=2$~bins (Kes 67, Kes 69,  Kes78, and W44) or 2.5~bins (W28) 
as shown in right panels in figure~\ref{fig:images}.
We call the \Ka\ line enhanced region in each SNR as the ``enhanced region'', while the nearby region is the ``reference region'', 
which are indicated by the dashed lines and solid lines, respectively. 
The enhanced and reference regions have typically $\sim1.2$ and $\sim0.6$~photons per bin, respectively.
We should note that the figures show rough distributions of the \Ka\ intensity due to the limited statistics.

We also made
 the continuum band images of the low and high energies, for comparison. Here, we did not use the events of the BI CCD (XIS\,1) because of poorer signal-to-noise ratios 
in the high-energy band. Figures~\ref{fig:images}a--\ref{fig:images}e show the XIS images of W28, Kes\,67, Kes\,69, Kes\,78, and W44, respectively. 
In the soft X-ray band (0.5--2.0~keV), we see clear emission associated with the SNRs, while in the hard X-ray band of 5.0--8.0 keV, no structure accompanied with the soft band is found. 
Thus, all these SNRs are soft X-ray sources with electron temperature of $\lesssim 0.8$~keV, which are confirmed in 
\cite{2012PASJ...64...81S}, \cite{2012A&A...541A.152B}, \cite{2016ApJ...818...63B}, and \cite{2012PASJ...64..141U}
for W28, Kes\,69, Kes\,78, and W44, respectively.
Kes 67 is also a soft X-ray source with an electron temperature of  $\sim0.4$~keV (section~\ref{ironline}). 

\begin{figure*}[tbh]
\figurenum{2}
\begin{center}
\includegraphics[width=5.5cm]{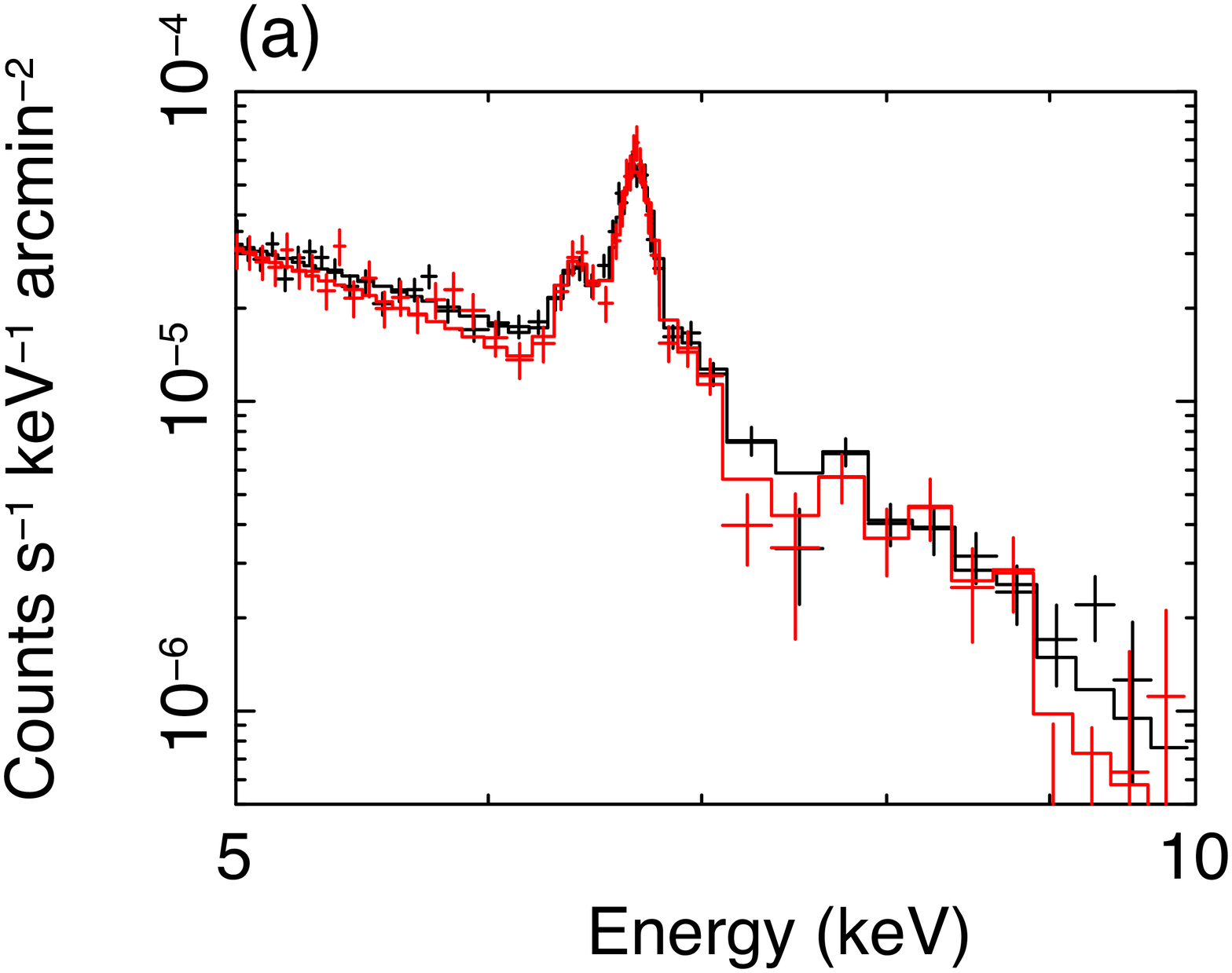}
\includegraphics[width=5.5cm]{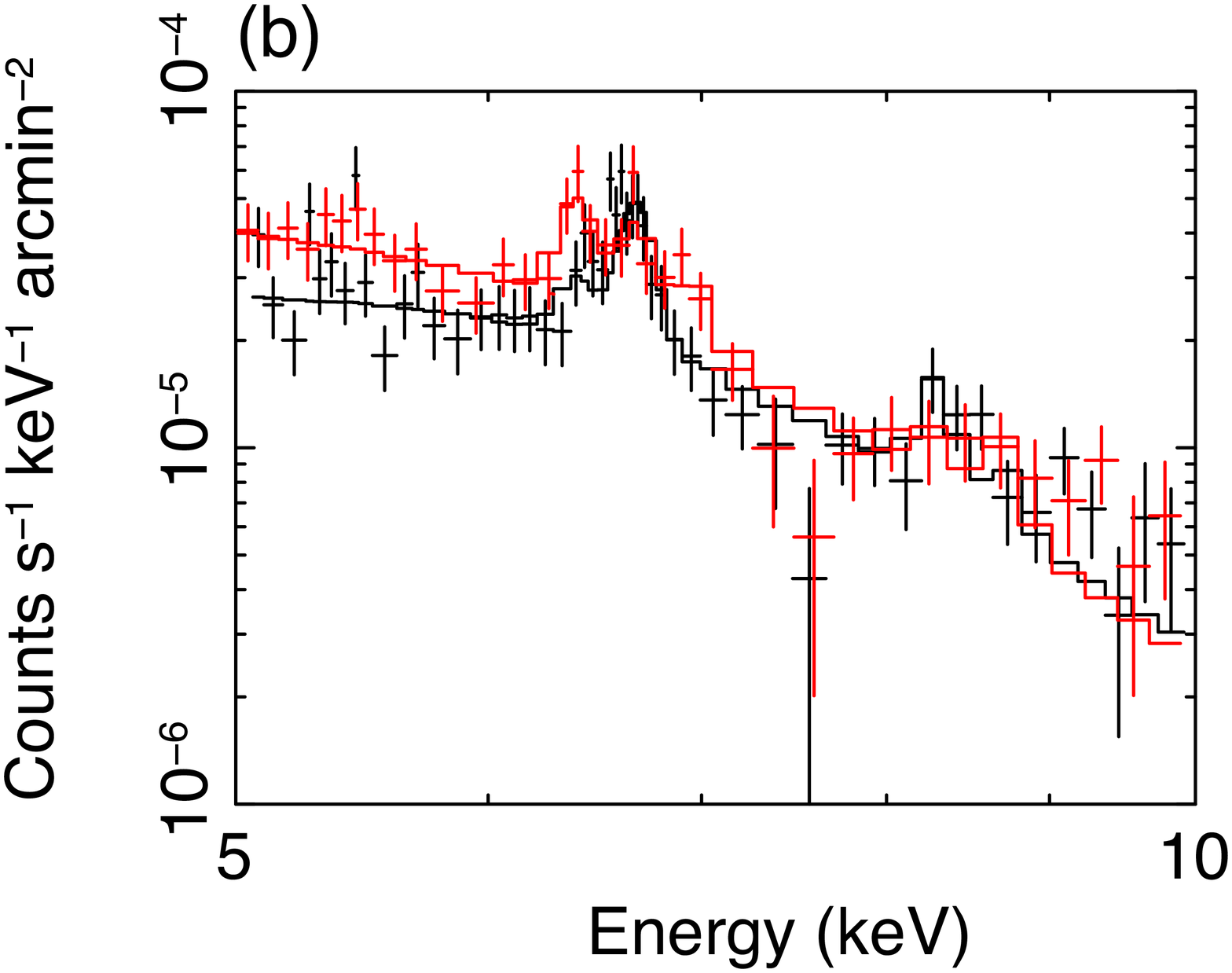}
\includegraphics[width=5.5cm]{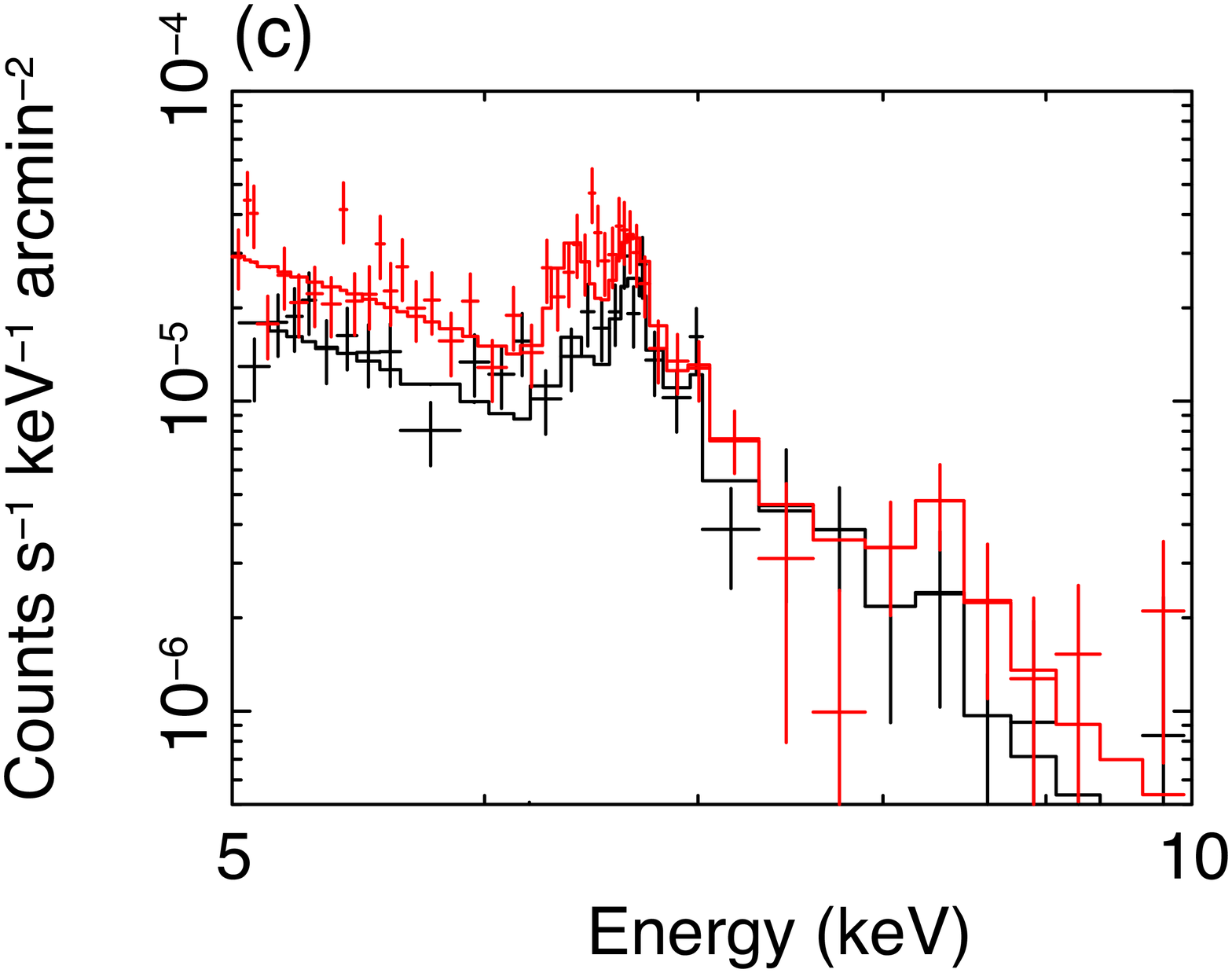}
\includegraphics[width=5.5cm]{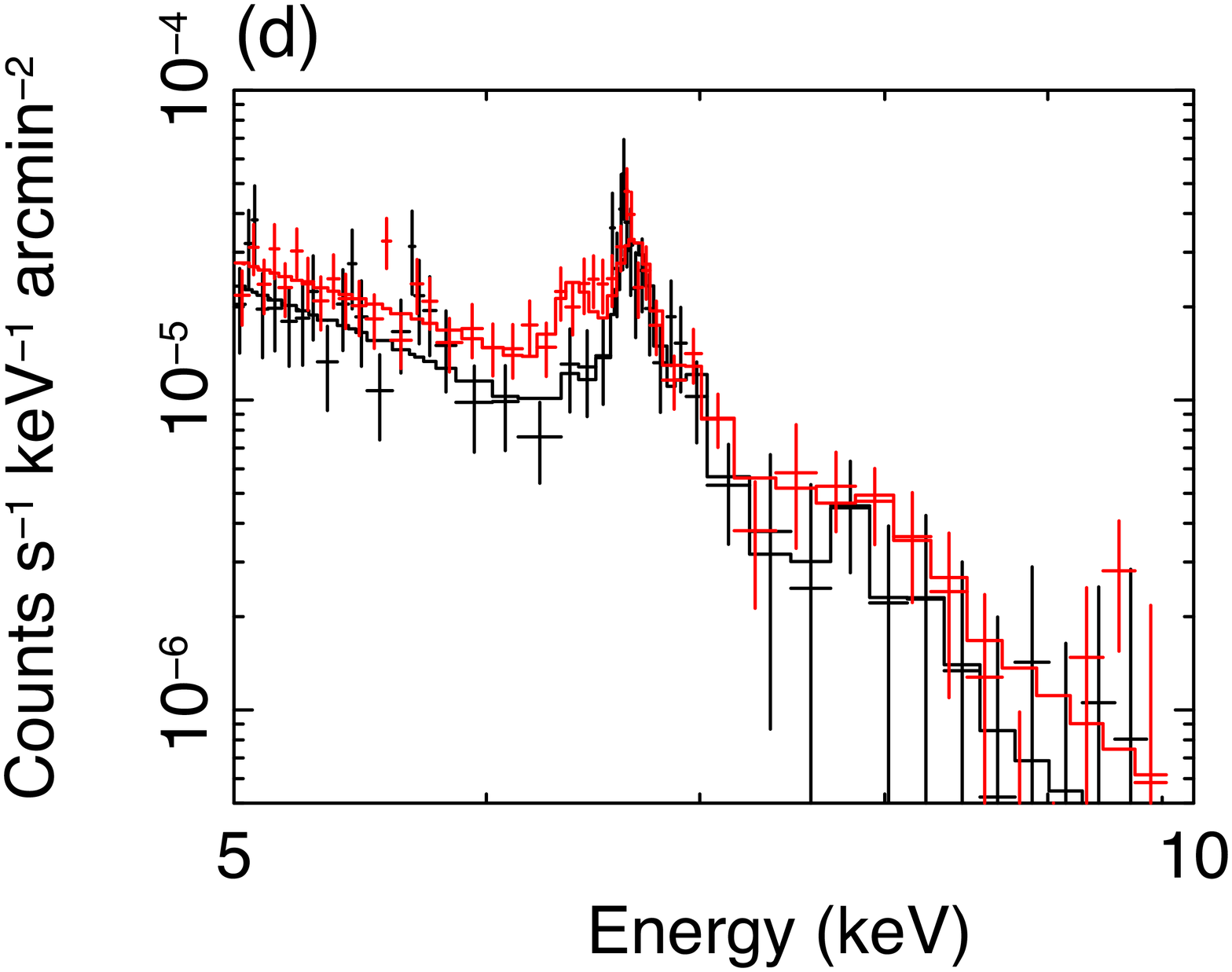}
\includegraphics[width=5.5cm]{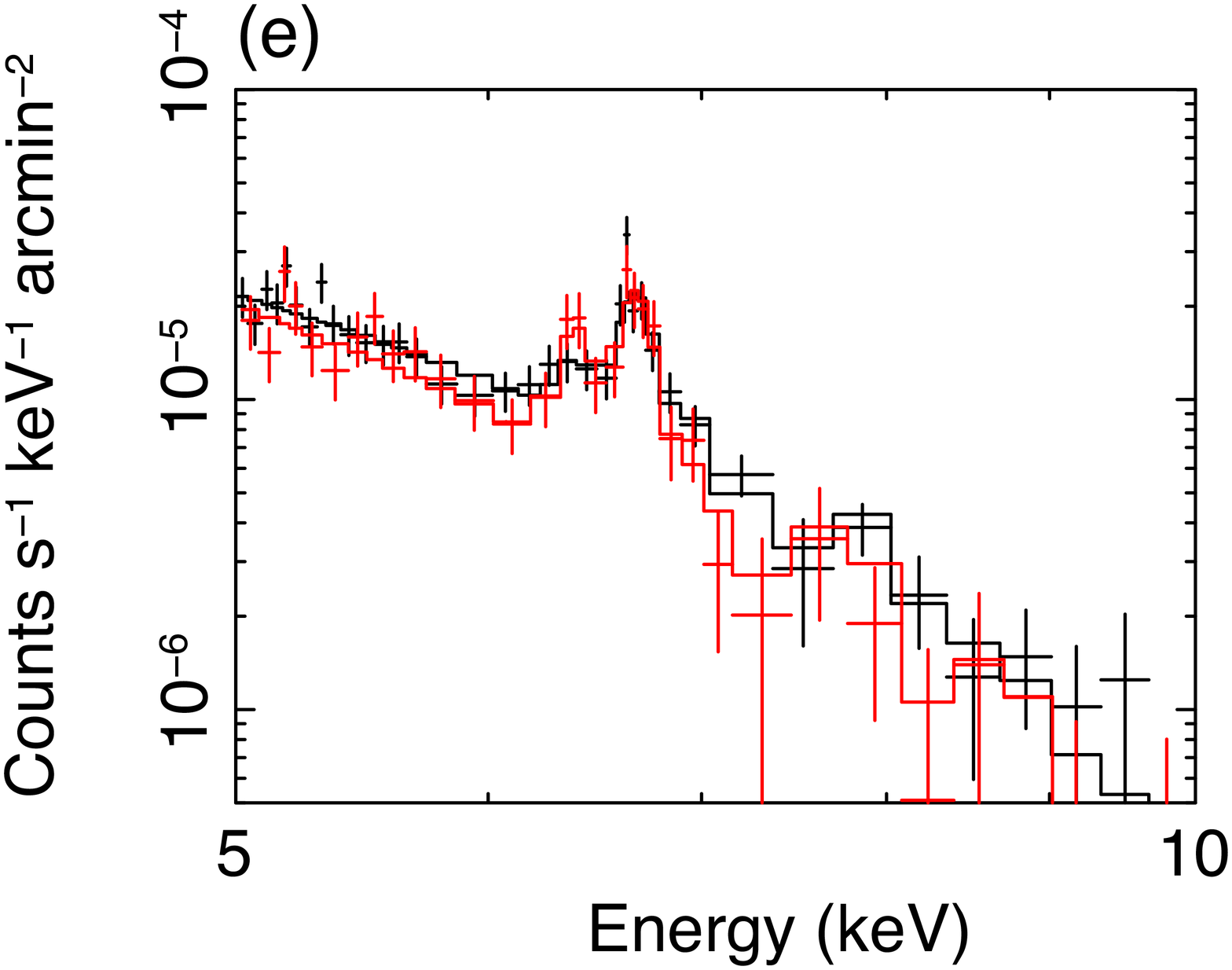}
\caption{
Spectra extracted from the \Ka\  enhanced (red) and reference (black) regions of 
(a) W28, (b) Kes\,67, (c) Kes\,69, (d) Kes\,78, and (e) W44.
The solid lines show the best-fit models consisting of a power law plus ten Gaussians (see text).  Errors are quoted at 1\,$\sigma$ confidence levels. 
\label{fig:2spec}
}
\end{center}
\end{figure*}

\floattable
\begin{deluxetable}{lcccccccccc}
\tablecaption{Intensities of the \Ka, \Hea, and \Lya\   lines and intensity ratios in the enhanced and reference regions. \label{tab:line}}
\tablecolumns{11}
\tablenum{2}
\tablewidth{0pt}
\tablehead{
\colhead{Object} &
\multicolumn2c{\Ka\  \tablenotemark{a,b}} &
\multicolumn2c{\Hea\  \tablenotemark{a,b}} &
\multicolumn2c{\Lya\  \tablenotemark{a,b}} &
\multicolumn2c{\Ka\ ratio \tablenotemark{b,c}} &
\multicolumn2c{\Lya\ ratio  \tablenotemark{b,d}} \\
\colhead{} &
\colhead{Enhanced} &
\colhead{Reference} &
\colhead{Enhanced} &
\colhead{Reference} &
\colhead{Enhanced} &
\colhead{Reference} &
\colhead{Enhanced} &
\colhead{Reference} &
\colhead{Enhanced} &
\colhead{Reference} 
}
\startdata
W28 & $3.14\pm0.43$ & $2.33\pm0.29$ & $9.80\pm0.53$ & $9.42\pm0.35$ & $1.52\pm0.43$ & $1.29\pm0.30$ & $0.32\pm0.05$ & $0.25\pm0.03$ & $0.16\pm0.04$ & $0.14\pm0.03$\\
Kes\,67 & $6.00\pm1.14$ & $2.39\pm0.73$ & $5.72\pm1.16$ & $7.35\pm0.84$ & $3.74\pm1.14$ & $<1.15$ &	$1.05\pm0.29$ & $0.33\pm0.11$ & $0.65\pm0.24$ & $<0.18$ \\
Kes\,69 & $4.35\pm0.74$ & $1.57\pm0.59$ & $5.44\pm0.77$ & $4.38\pm0.65$ & $0.76\pm0.69$ & $1.24\pm0.65$ &$0.80\pm0.14$ & $0.36\pm0.14$ & $0.14\pm0.13$ & $0.28\pm0.15$ \\
Kes\,78 & $2.62\pm0.56$ & $0.81\pm0.69$ & $5.19\pm0.61$ & $5.66\pm0.80$ & $1.33\pm0.61$ & $1.62\pm0.72$ & $0.50\pm0.12$ & $0.14\pm0.12$ & $0.26\pm0.12$ & $0.29\pm0.13$ \\
W44 & $2.18\pm0.55$ & $0.92\pm0.32$ & $3.66\pm0.57$ & $3.47\pm0.35$ & $0.96\pm0.60$ & $0.91\pm0.33$ &$0.60\pm0.18$ & $0.27\pm0.10$ & $0.26\pm0.17$ & $0.26\pm0.10$ \\
\enddata
\tablenotetext{a}{Units are $10^{-8}$~photons~cm$^{-2}$~s$^{-1}$~arcmin$^{-2}$.}
\tablenotetext{b}{Errors are quoted at the 1$\sigma$ level.}
\tablenotetext{c}{Intensity ratio of \Ka/\Hea.}
\tablenotetext{d}{Intensity ratio of \Lya/\Hea.}
\end{deluxetable}
%\vspace{5mm}

In W28, Kes\,69, Kes\,78, and W44, some point-like or slightly extended sources are found (figures~\ref{fig:images}a, \ref{fig:images}c, \ref{fig:images}d, and \ref{fig:images}e). The point source in the south-west of W28 is reported by 
\cite{2017ApJ...839...59P}; the object, CXOU J175857.55--233400.3, 
is likely a cataclysmic variable  or a quiescent low-mass X-ray binary. Two 
point-like sources located in the north and the west of Kes\,69 are reported by  \cite{2012A&A...541A.152B}. 
However no further information on these sources is available. The bright point source in 
the north of Kes\,78 is 2XMM\,J185114.3--000004. \cite{2016ApJ...818...63B} suggested that 
this object can be a supergiant fast X-ray transient. In the central region of W44, we see two extended sources. The northern source is reported by 
\cite{2012PASJ...64..141U}. The authors claimed that this hard X-ray emission can be due to a synchrotron source.  However, thanks to deep follow-up observations, our quick analysis
 revealed a hint of an emission line at $\sim6.15$~keV from this hard diffuse source. Thus this source is likely a new identified cluster of galaxies behind the Galactic plane with the
 redshift of $z\sim0.09$. 
 The southern source is a PWN associated with PSR B1853+01 \citep{1996ApJ...464L.165F}, and was detected in X-rays by ASCA \citep{Harrus:1996di}. 

In the right panels of figure~\ref{fig:images}, we see that the morphology of the \Ka\ line emission is clearly different from the soft-band emission. 
The NXB-subtracted spectra of the enhanced and reference regions for the individual SNRs are shown in figure~\ref{fig:2spec}. 
We found excess in the 7--8 keV band, which would come from the \Heb, -He$\gamma$, \Lyb, -Ly$\gamma$, \NiKa, and \NiHea\ lines associated with the GRXE \cite[e.g.,][]{2016ApJ...833..268N}. We, then, added six 
Gaussians to the model, and set the centroids to the values found in AtomDB 3.0.2  (http://www.atomdb.org/). The best-fit fluxes of the \Ka, \Hea, and \Lya\ lines are shown in table~\ref{tab:line}, 
and the best-fit models are shown in figure~\ref{fig:2spec}. 

We found that the \Ka\ line intensity in the enhanced region is stronger than that in the reference region for each SNR. The significance levels are $1.6\,\sigma$, $2.7\,\sigma$, $2.9\,\sigma$, $2.0\,\sigma$, and $2.0\,\sigma$ for W28, Kes\,67, Kes\,69, Kes\,78, and W44, respectively. On the other hand, the \Hea\ line intensity in the enhanced region is consistent with the reference region within the $1\,\sigma$ error range in each SNR. 
The \Lya\ line intensities in the enhanced regions are also consistent with the reference regions within the $1\,\sigma$ error range in all the SNRs except Kes\,67.  In Kes\,67, the \Lya\ line intensity of the enhanced region is stronger than that of the reference region. However, the poor statistics prevents detailed investigation of this possible \Lya\ enhancement. 

\floattable
\begin{deluxetable}{lccccc}
\tablecaption{Intensities of the \Ka, \Hea, and \Lya\  lines and intensity ratios of the integrated spectra of the enhanced  and reference regions and the GRXE. \label{tab:integ}}
\tablecolumns{6}
\tablenum{3}
\tablewidth{0pt}
\tablehead{
\colhead{Region} &
\multicolumn3c{Line intensity\tablenotemark{a,b}} &
\multicolumn2c{Intensity ratio\tablenotemark{b}}\\
\colhead{} &
\colhead{\Ka\ } &
\colhead{\Hea\ } &
\colhead{\Lya\  } &
\colhead{\Ka/\Hea\ } &
\colhead{\Lya/\Hea\ } 
}
\startdata
Enhanced & $3.37\pm0.54$ & $5.73\pm0.56$ & $1.47\pm0.53$ & $0.59\pm0.11$ & $0.26\pm0.10$ \\
Reference & $1.60\pm0.42$ & $5.83\pm0.48$ & $1.00\pm0.43$ & $0.27\pm0.08$ & $0.17\pm0.08$ \\ \hline
GRXE \tablenotemark{c} & $1.6\pm0.2$ & $6.0\pm0.3$  &  $1.0\pm0.2$ & $0.27\pm0.03$ & $0.17\pm0.03$ \\
\enddata
\tablenotetext{a}{Units are $10^{-8}$~photons~cm$^{-2}$~s$^{-1}$~arcmin$^{-2}$.}
\tablenotetext{b}{Errors are quoted at 90\% confidence levels.}
\tablenotetext{c}{$10\degr<|l|<30\degr$ and $|b|<1\degr$ \citep{2016ApJ...833..268N}.}
\end{deluxetable}
%\vspace{5mm}

\begin{figure}[tb]
\figurenum{3}
\begin{center}
\includegraphics[width=7cm]{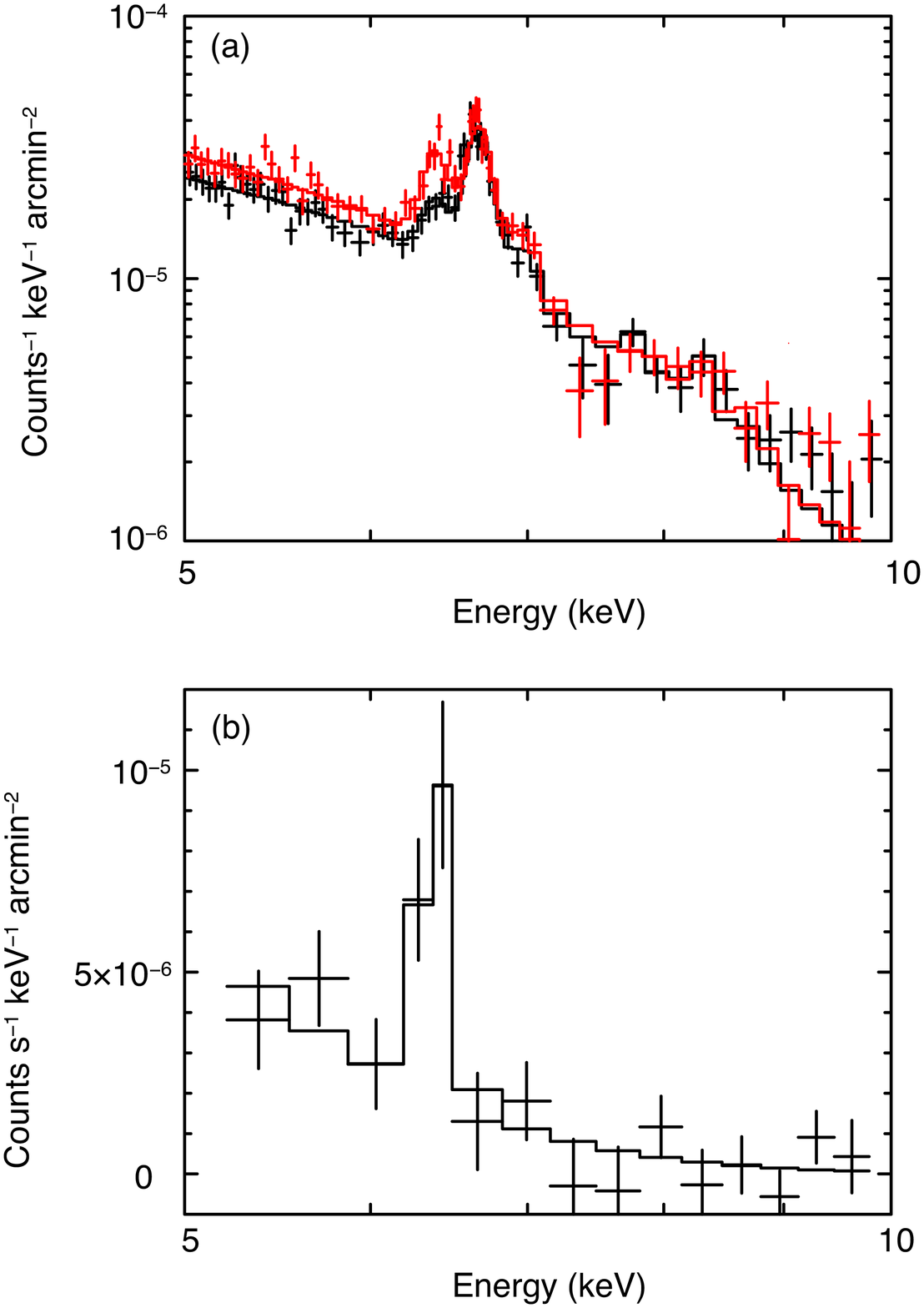}
\caption{(a) X-ray spectra extracted from the enhanced region (red) and the reference region (black). 
The red and black lines show the best-fit models consisting of a power law plus ten Gaussians (see text). 
(b) X-ray spectrum of the enhanced emission, which is obtained by subtracting the spectrum of the reference region from that of the enhanced region (see text).  Errors are quoted at 1\,$\sigma$ confidence levels. 
\label{fig:2spectra}
}
\end{center}
\end{figure}

In order to determine whether the \Ka\ line from the ensemble of the SNRs is statistically significant,
we made the stacked spectra for the enhanced and reference regions of the individual  SNRs, as are shown in figure~\ref{fig:2spectra}a. 
The two spectra were fitted with the same phenomenological model used for the  individual SNRs. The best-fit curves are given in figure~\ref{fig:2spectra}a, 
while the best fit parameters of the \Ka, \Hea\ and \Lya\ intensities are listed in table~\ref{tab:integ}. 
The flux ratio of \Ka\ between the enhanced and reference regions is $2.1\pm{0.6}$, 
while those of \Hea\ and \Lya\ are $0.98\pm0.13$ and $1.5\pm0.8$, respectively (at 90\% confidence levels).  
Thus, a clear enhancement of the \Ka\ line from the surrounding reference regions is found with more than 4.3\,$\sigma$ confidence,
while no enhancement of  \Lya\ is found.

We, then, subtracted the stacked spectrum of the reference regions from that of the enhanced regions. 
Since the GRXE dominates the reference and enhanced spectra, this subtraction should 
be based on a reliable GRXE estimation in both the spectra.  The relevant GRXE may be often
contaminated by the stray light of nearby bright sources \citep{1989Natur.339..603K}.
This contamination in an iron K-shell line is smaller than that in the continuum (e.g., the 5--8 keV band),
because the equivalent width (EW; the line flux divided by the continuum flux) of the iron K-shell line in the Galactic bright sources is typically $\lesssim$100 eV,
which is far smaller than that of the GRXE \citep[e.g.,][]{1989Natur.339..603K, 2006MNRAS.373L..11R}. 
Thus, assuming that the \Hea\ line flux is a good indicator for the flux of the GRXE, we fine-tuned the positional difference of the GRXE flux between the enhanced and reference regions, using the \Hea\ line intensity (see \citealt{2015ApJ...807L..10N}); 
we made the spectrum of the enhanced emission taking account of the difference of the \Hea\ line intensity by 1.7\%.  
The result is given in figure~\ref{fig:2spectra}b.
We fit the obtained spectrum with a power law plus a Gaussian at 6.40~keV. 
The EW and photon index are measured to be $1.0^{+0.7}_{-0.4}$~keV and $4.1^{+3.8}_{-2.4}$, 
respectively (at 90\% confidence levels). 
For a conservative estimate of the lower limit of the \Ka\ EW,  we took account of the 90\% confidence error of the \Hea\ line intensity. 
Then, we obtained that the \Ka\ EW is at least $\sim400$~eV. 
 Instead of the analysis of the stacked spectra, we tried subtracting the reference spectrum from the enhanced spectrum for each SNR and  fitting the five spectra simultaneously;  the result is consistent with the analysis of the stacked spectra.

\begin{figure}[tb]
\figurenum{4}
\begin{center}
\includegraphics[width=7cm]{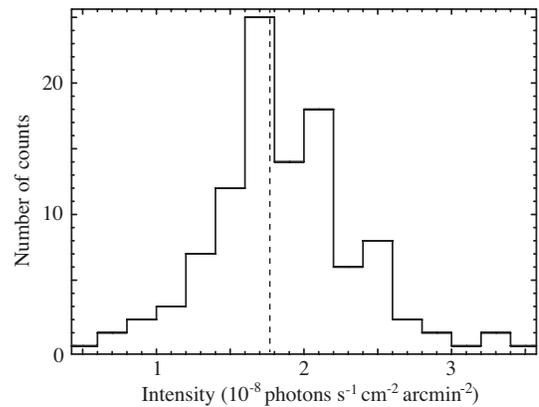}
\caption{
Distribution of the enhanced \Ka\ intensities obtained from 100 trials of the simulations. The distribution has a mean value and a standard deviation of $1.8\times10^{-8}$ and $0.4\times10^{-8}$~photons~s$^{-1}$~cm$^{-2}$~arcmin$^{-2}$, respectively. The dashed line indicates the best-fit value ($1.77\times10^{-8}$~photons~s$^{-1}$~cm$^{-2}$~arcmin$^{-2}$) obtained with the actual observation data (table~\ref{tab:integ}).
\label{fig:simu}
}
\end{center}
\end{figure}

Since the above stacking process might be a subject of systematic errors, we estimated the possible systematic 
errors based on simulations in this paragraph. 
We simulated the spectra from the enhanced and background regions of each SNR, and evaluated the \Ka\ intensities by performing the same stacking analysis as we did for the observational data. In this simulation process, we assumed  the exposure time as the actual observations. After 100 trials of the simulations, the distribution of the \Ka\ intensities obtained has a mean of $1.8\times 10^{-8}$ and a standard deviation of $0.4\times 10^{-8}$~photons~s$^{-1}$~cm$^{-2}$~arcmin$^{-2}$, which are almost the same as the enhanced \Ka\ line intensity and the statistical error obtained from the observational data. The distribution is shown in figure~\ref{fig:simu}. We furthermore performed the same simulation but with 100 times longer exposure assumed. Then, the dispersion became $0.05\times 10^{-8}$~photons~s$^{-1}$~cm$^{-2}$~arcmin$^{-2}$, one order of magnitude smaller than that calculated with the actual exposure. Therefore, we conclude that the stacking process introduces no significant systematic errors which dominate over the statistical errors.

\section{Discussion}
\subsection{Origin of the \Ka\ enhancement} \label{sec:origin} % sec 4.1
In the region of $6\degr \lesssim l \lesssim 40\degr, |b| < 1\degr$, we have found the enhanced \Ka\ emission  from five SNRs, W28, Kes\,67, Kes\,69, Kes\,78 and W44. In the same region, \cite{2014PASJ...66..124S, 2016PASJ...68S...8S} also found \Ka-enhancements from Kes\,79 and 3C\,391.
The origin of the \Ka\ emission would be the same.
In comparison with the \Hea\ and \Lya\ lines, we discuss the origin of the \Ka\ line emission from the SNRs. 

\subsubsection{Fluctuations of the GRXE?} % sec 4.1.1
The X-ray emission in the $> 5$~keV band in each SNR is dominated by the GRXE.  Is the enhanced \Ka\ emission simply a statistical fluctuation of the GRXE?  To answer this question, we list the mean iron K-shell line flux of the GRXE ($|l|= 10\degr$--30$\degr$) in table~\ref{tab:integ}, following \cite{2016ApJ...833..268N}. 
In table~\ref{tab:integ}, the line fluxes of \Ka, \Hea\ and \Lya\ in the reference region agree to those of the
mean GRXE.  Also the \Hea\ and \Lya\ line fluxes in the enhanced region agree with those of the mean GRXE. 
As is noted in section~\ref{sec:image_spec}, the \Hea\ line flux is a more reliable indicator of the 
GRXE flux than the 5--8 keV flux. 
We, therefore, use the flux ratios of \Ka/\Hea\  instead of the EW 
for estimation of the GRXE contribution.
The flux ratios of the iron K-shell lines between the enhanced and reference region are 
also listed in table~\ref{tab:integ}. From table~\ref{tab:integ},  
we conclude that the \Ka\ line enhancements are not due to statistical fluctuation of the GRXE, but associated with the SNRs.

\subsubsection{Thermal plasma origin?} % sec 4.1.2
The iron K-shell line emission has been detected in many SNRs.  
Most of the iron is of the ejecta origin, and hence is located around the center of a supernova explosion. 
The iron-rich ejecta are heated by the reverse shock.  
Therefore, the ionization state is often lower than helium-like iron in young  or middle-aged SNRs, 
which is called an ionizing plasma (IP).  
The iron K$\alpha$ centroids  depend on the ionization age, and hence some young  SNRs show a very low ionization state of iron with the line centroids of $\sim6.4$~keV \citep[e.g., RCW\,86, SN\,1006;][]{2008PASJ...60S.141Y, Yamaguchi:2008wc}. 
One may argue that the observed 6.40~keV line in the soft X-ray SNRs of our sample is also due to IP. 
However, with the following two reasons, we interpret that the 6.40 keV enhancement would not be of the IP origin.

One is that the morphology of the 6.40~keV emission is clearly different from that of the soft-energy band, the distribution of the SNR plasma.
In particular, in Kes\,67 and W44, the distributions of the 6.40~keV emission are not center-filled, like the image of the soft-energy band, but have shell-like structures.
This may indicate that the 6.40~keV emission does not arise from the SNR plasma, 
although the different morphologies do not necessarily mean a different origin.

The other reason is that all the SNRs with the 6.40 keV line enhancements interact with MCs \citep{1997ApJ...489..143C, Tian:2007cl, 2009ApJ...691..516Z, 1998AJ....116.1323K, 2005ApJ...627..803H}, which indicates that these SNRs are likely core-collapse SNRs.
\cite{2014ApJ...785L..27Y} reported the iron K$\alpha$ centroids of  core-collapse SNRs are significantly higher than 6.40~keV. This is in a sharp contrast to  Type Ia SNRs,
 which show lower energy of $< 6.5$~keV.  
Since the ionization or excitation rate of  iron K-shell electrons decreases very rapidly below the electron temperature of 1~keV,
the iron K-shell line should be detected in SNRs with the electron temperature higher than $\sim1$~keV \citep{2014ApJ...785L..27Y}.
In other words, it is almost impossible for the SNRs in our sample to emit a detectable iron K-shell line because of the low electron temperature.

\subsubsection{Ionization by X-rays or cosmic-ray particles?}  % sec 4.1.3
Another scenario is that a bright X-ray source is located near 
each SNR and X-rays from the source irradiate the gas surrounding the SNR 
to produce the \Ka\ line. Assuming a typical value for the hydrogen column density (of the ambient gas) $N_{\rm H}=3\times10^{22}$~cm$^{-2}$ and a typical \Ka\ enhanced area of $10^2$--$15^2$~arcmin$^{2}$, the source luminosity should be 
$L_{\rm X} \sim 1 \times10^{38} (D/100~{\rm pc})^2$~erg~s$^{-1}$ \citep{Sunyaev:1998kz}, where $D$ is the distance of the \Ka\ line emitting region from the irradiating source. 
The required luminosity is close to the Eddington luminosity of a neutron star $L_{\rm Edd} = 2 \times10^{38}$~erg~s$^{-1}$. Such a bright source is not located nearby
 the SNRs. Thus the X-ray irradiation scenario is also difficult.
 
The most plausible scenario is inner-shell ionization of iron atoms in the surrounding gas by CRs accelerated in the SNRs. The EW of the 
\Ka\ line depends on the irradiating particle and the iron abundance. Here we assume the iron abundance of 1 solar. In the case of the proton
 bombardment, the EW is around or higher than $\sim 500$~eV, while in the case of the electron bombardment, the EW is below 400~eV due to strong
 bremsstrahlung, regardless of the spectral index of the particles \citep{2011PASJ...63..535D}. The lower limit of the measured EW of 400~eV rejects the electron origin. The \Ka\ line enhancements in the five SNRs would be due to interactions between LECRp and the ambient cold gas.

\subsection{Implication for the origin of the GRXE}  % sec 4.2
\cite{2016ApJ...833..268N} reported that the mean spectrum of the GRXE shows significant enhancements of the \Ka, \Hea, and \Lya\ lines from the assembly of the known point sources, cataclysmic variables and X-ray active binary stars. On the other hand, in general, the point sources explain the continuum of the GRXE spectrum.  Therefore, the other candidates emitting iron lines are required. 
As is discussed in section~\ref{sec:origin}, the excess \Ka\ intensity over the mean GRXE is found in  our SNR samples.
Since the hard X-ray fluxes (5--8~keV) of the SNRs  are negligibly small due to their low temperature of 
$\lesssim1$~keV, these SNRs may not contribute to the GRXE in the 5--8~keV band.
Therefore, the candidate sources of the \Ka\ excess in the GRXE are LECRp colliding with cold gas in SNRs. 
The \Ka\ line excess would be more common among SNRs that are faint in hard X-ray or are not detected in X-ray.

Such SNRs may also have some contribution to the excess of the \Hea\ and \Lya\ lines in the GRXE above the predicted point-source contribution. 
Some of the SNRs that show the \Ka-enhanced emission are known to be have recombining plasma (RP) for silicon and sulfur \citep[W28, W44, and 3C\,391;][]{2012PASJ...64...81S, 2012PASJ...64..141U, 2014PASJ...66..124S}.  
Since LECRp can ionize helium-like silicon and sulfur in the hot plasma of SNRs to hydrogen-like ions, they  may contribute to production of RP.  

LECRp may also generate helium-like and hydrogen-like ions of iron. 
We roughly estimate ionization rate of helium-like iron for the cases where iron is ionized by LECRp or thermal electrons in the plasma with $kT\sim1$~keV. The ionization rate is proportional to the LECRp/electron density, velocity, and the cross section. Number density of the MeV protons obtained in this study is $n_{\rm p}\sim10^{-5}$~cm$^{-3}$ (equivalent to the energy density of 100~eV~cm$^{-3}$; see section~\ref{sec:density}). In the thermal plasma,  electrons with energy above the K-shell ionization energy of iron are $\sim0.01$\% of the total; assuming the total electron density of 1~cm$^{-3}$, we obtain $n_{\rm e}\sim10^{-4}$~cm$^{-3}$. Ionization cross section of helium-like iron for electrons with the kinetic energy of $\sim$10~keV is $\sigma_e \sim10^{-22}$~cm$^{-2}$, which is the same in the order of magnitude as that for LECRp ($\sigma_{\rm p}$) with the kinetic energy of $\sim$10~MeV \citep{2000A&A...357..716K}. Also velocities of electrons ($v_{\rm e}$) and LECRp ($v_{\rm p}$) are similar to each other ($\sim0.2c$). Thus, the ratio of the ionization rate between the cases of LECRp and thermal electrons are $(n_{\rm p} v_{\rm p} \sigma_{\rm p})/(n_{\rm e} v_{\rm e} \sigma_{\rm e}) \sim0.1$. The LECRp can contribute to production of RP.

For example,
a soft X-ray SNR, IC443 has RP emitting the \Hea\ and \Lya\ lines \citep{2014ApJ...784...74O}. 
Although the mechanism is under discussion, LECRp may contribute. 
At present there are limited samples of the hard X-ray faint SNRs with the \Hea\ and \Lya\ excess,  
but if the number of such samples increases by further observations, we will possibly be able to evaluate their contribution to the GRXE.

\subsection{Origin of the MeV protons}  % sec 4.3
\label{sec:CO}
The \Ka\ line emission is expected to be associated with dense gas such as MCs and \ion{H}{1} clouds. Therefore, we compared the distribution of the
\Ka\ line emission with that of \ion{H}{1} clouds for W28 and Kes\,67, and with that of the CO intensity for Kes\,69, Kes\,78, and W44. 
All the five SNRs have been well studied by radio observations. The velocities (LSR) where the individual SNRs are associated with molecular clouds were identified by molecular line broadening (all the SNRs), 1720\,MHz OH maser  (W28, Kes\,69, Kes\,78, and W44),   and a high line intensity ratio $^{12}$CO$ (J = 2-1)/^{12}$CO$ (J = 1-0)$ (Kes\,67 and W44) as well as morphological association \citep[][and references therein; \citealt{2011ApJ...743....4Z}]{2010ApJ...712.1147J}. The velocities of the \ion{H}{1} and CO intensity maps in figure~\ref{fig:images} are consistent with those velocities.
In W28, Kes\,67, Kes\,69, and W44, we found association between the \Ka\ line and dense gas clouds 
interacting with each SNR (figure~\ref{fig:images}).
The Larmor radius of LECRp is extremely short ($\sim10^{11}$~cm at 10~MeV in the magnetic field of 1~$\mu$G). 
Therefore, the \Ka\ line should be emitted just near the LECRp acceleration sites. 

 \cite{2015ApJ...807L..10N} reported that diffuse  \Ka\  line emission was discovered in a region 3\degr~eastern region of the Galactic center. The authors claimed that the  \Ka\  line emission is generated by LECRp. Because of the short diffusion length of LECRp, the protons should be produced in situ. There is no SNR in the region. The \Ka\  line emission is considered to be associated with a giant molecular cloud, Bania's Clump~2 exhibiting a large velocity dispersion $\sim100$~km~s$^{-1}$ \citep{1977ApJ...216..381B, 2010PASJ...62.1307T}. The authors claimed that the LECRp can be produced by stochastic acceleration via alfv\'enic turbulence. In our case, on the other hand, there is no molecular cloud with such a large velocity dispersion, and hence the same scenario cannot be applied. The most possible scenario is that the MeV protons come from the SNRs.
The spatial association between the \Ka\ line emission and the dense clouds suggests that the MeV protons are produced by shocks interacting with the dense clouds. 

\begin{figure*}
\figurenum{5}
\begin{center}
\includegraphics[width=10cm]{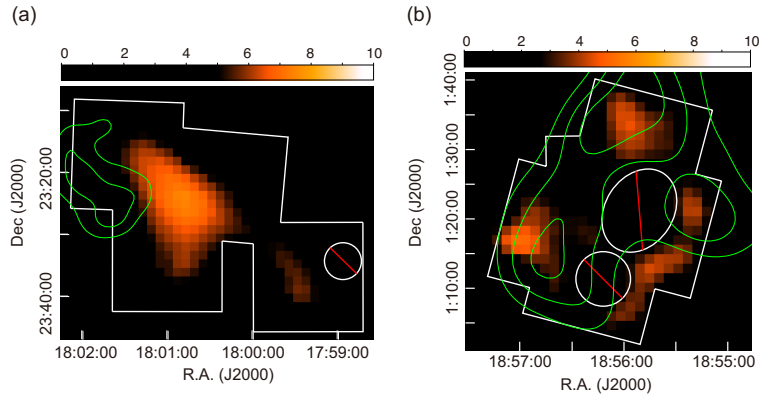}
\caption{Distributions of the \Ka\  band (same as the right panels of figures~\ref{fig:images}a and \ref{fig:images}e) overlaid with contours of (a) TeV gamma-ray emission  for W28 \citep{2008A&A...481..401A} and (b) gamma-ray emission in the 2--10~GeV band which is observed by {\it Fermi} LAT \citep{2010Sci...327.1103A} for W44.}
\label{fig:TeV_GeV}  
\end{center}
\end{figure*}

Since the LECRp would be a seed particle to the high energy acceleration up to the GeV and TeV energy, we compared the distribution of the \Ka\ line with those of the high energy  (GeV and TeV) gamma rays.  
The gamma-ray results are available only for W28 and W44; figures~\ref{fig:TeV_GeV}a and \ref{fig:TeV_GeV}b present the TeV and GeV gamma-ray contours  of W28 and W44, respectively, overlaid on the \Ka\ line band images. 
The peak of the TeV gamma-ray distribution of W28 lies to the east by $\sim 15'$ from the \Ka\ line emission and is located on the FOV edge of {\it Suzaku}, where the observing efficiency is poor. The gamma-ray emission is positionally associated with MCs \citep{2008A&A...481..401A}. 
Since the LECRp cannot penetrate into the MC, this image is not inconsistent with that the MeV protons are the origin of TeV protons. 
The \Ka\ line emission far from the TeV emission would be due to low $N_{\rm H}$ region ($\sim$ several times of $10^{22}$~cm$^{-2}$) of neutral gas. 
The GeV gamma rays in W44 are interpreted to be produced via interactions between GeV protons and the MCs \citep{2010Sci...327.1103A,2010A&A...516L..11G}. 
The \Ka\ line could be generated from the low-energy protons accelerated in the GeV proton region. 
The GeV gamma-ray and \Ka\ line regions well overlap, in spite of the extremely small path length of MeV protons (the \Ka\ region) compared to the large one of GeV protons (the GeV gamma-ray region).  This would indicate that the \Ka\ region is located just in front of or behind the gamma-ray region.

\subsection{Energy density of protons}% sec 4.4
\label{sec:density}
In the same manner as in \cite{2015ApJ...807L..10N}, we estimated the energy density of LECRp. The hydrogen column densities of the MCs associated
 with the SNRs range from $\sim5\times10^{21}$ to $\sim3\times10^{22}$~cm$^{-2}$ \citep{2008AIPC.1085..104F, Tian:2007cl, 2009ApJ...691..516Z, 2011ApJ...743....4Z, 2013ApJ...768..179Y}.
  Although the \Ka\ enhancements are not always consistent with the MCs (see section~\ref{sec:CO}), we assume these values as the upper limit of the
 hydrogen column densities of the ambient gas associated with the SNRs, and obtain the lower limit of the proton energy density. If we assume a
 mono energetic distribution, the lower limit of the proton energy density at 10 MeV, where the cross section has a peak of $1.3\times10^{-26}$~cm$^{2}$~hydrogen-atom$^{-1}$ in solar abundance \citep{2012A&A...546A..88T, 2011PASJ...63..535D, 2003ApJ...591.1220L}, is estimated to be $\sim10$--$100$~eV~cm$^{-3}$.  
 This value is comparable to or one order magnitude higher than the extrapolation of the high-energy CRs (see section~\ref{sec:intro}).

\subsection{Future works}% sec 4.5
Among the active X-ray astronomy satellites, {\it XMM-Newton} can be used to separate the \Ka\ line \citep[e.g.][]{2012A&A...546A..88T}. We plan combination analysis of {\it Suzaku} and {\it XMM-Newton}  to strengthen the LECR scenario.  Moreover, in  future X-ray missions, we will be able to perform spectroscopy with a high-energy resolution ($\sim {\rm eV}$) and to measure the \Ka\ line much more sensitively.       

\section{Conclusions}
We found the \Ka\ line from the five SNRs, W28, Kes\,67, Kes\,69, Kes\,78, and W44 among seven samples located in the region of $6\degr < l < 40\degr$ and $|b| < 1\degr$ observed by {\it Suzaku}. 
The X-ray spectrum of the \Ka\ excess has a large EW ($>400$~eV) and is well explained by   interactions between LECRp, or MeV protons, and the ambient cold gas. The energy density of the LECRp is estimated to be at least 10--100~eV~cm$^{-3}$.

\acknowledgments

We thank all the members of the {\it Suzaku} team. 
We are grateful to Ping Zhou for providing the CO data shown in figure~\ref{fig:images}d. 
K.K.N. is supported by Research Fellowships of JSPS for Young Scientists. 
This work was supported by JSPS and MEXT KAKENHI Grant Numbers JP16J00548 (K.K.N.), JP17K14289 (M.N.), JP25109004 (T.T.), and JP26800102 (H.U.).

\end{document}